\documentclass[aps,prl,groupedaddress]{revtex4}
\usepackage{graphicx}
\usepackage{dcolumn}
\usepackage{bm}
\input epsf
\epsfclipon
\begin{document}

\title{Self-organized criticality in a model of collective bank bankruptcies}
\author{Agata Aleksiejuk (a,b), Janusz A. Ho\l yst (a,b) and Gueorgi Kossinets (c)}
\affiliation{(a) Faculty of Physics, Warsaw University of Technology, Koszykowa 75, PL-00-662 Warsaw, Poland \\ 
(b) Institute for Economics and Traffic, Dresden University of
Technology, Andreas-Schubert-Stra\ss e 23, D-01062 Dresden,
Germany \\ (c) Department of Sociology, Columbia University, 413
Fayerweather Hall, 1180 Amsterdam Ave., Mail Code 2551, New York,
NY 10027, USA 
\\
e-mail: jholyst@if.pw.edu.pl}

\bigskip

\begin{abstract}
The question we address here is of whether phenomena of collective
bankruptcies are related to self-organized criticality. In order
to answer it we propose a simple model of banking networks based
on the random directed percolation. We study effects of one bank
failure on the nucleation of contagion phase in a financial
market. We recognize the power law distribution of contagion sizes
in 3d- and 4d-networks as an indicator of SOC behavior. The SOC
dynamics was not detected in 2d-lattices. The difference between
2d- and 3d- or 4d-systems is explained due to the percolation
theory.

\end{abstract}

\maketitle

\section{Introduction}
In recent years, physicists have intensively studied economic and
social systems composed of many mutually interacting parts
\cite{a1,a2,a3,a4}. The methods borrowed from physics provided a
better understanding of such phenomena like crashes in a stock
market \cite{a6}, occurrence of extremely large market shares
\cite{a7}, and emergence of totalitarianism in democratic
societies \cite{a8,a9,a10,a11}. Like many other economic
phenomena, collective bankruptcies arise from the complex nature
of financial and capital markets \cite{a12,a15,a14,a13}. Here we
focus on one of the most extreme examples of systemic failure,
namely bank bankruptcies. We use banking terminology following our
previous studies on interbank markets \cite{a12}, but the proposed
model is general enough to describe many other types of collective
failures.

As emphasized in \cite{a13}, the history of modern banking is full
of examples of systemic failures at both moderate and large
scales. Perhaps most of the readers are familiar with anecdotal
accounts of the Great American Depression of 1929, or those of the
1997 East Asian crisis. On the other hand, very few people noticed
financial troubles of Bank \'{S}l\c{a}ski a few years ago in
Poland. What is the reason behind the fact that one bankruptcy
triggers off an avalanche of further failures while other does
not?

\section{Banking reality and network model}
There are two main characteristics describing financial weakness
of a bank: liquidity and solvency. Liquidity is the ability to pay
short-term commitments on time, whereas solvency encompasses all
commitments. Loss of solvency is a sufficient condition for
bankruptcy and loss of liquidity is its necessary condition,
respectively. Banks face shortfalls and surpluses of money
resulting from imbalances between revenues and expenses. This
additional or lacking money may be balanced via interbank lending.
In this way interbank lending prevents banks from the loss of
liquidity.

At present, interbank market is an important element of money
market. In most of interbank markets the mean transaction period
is shorter than one month and still shortening. Moreover, the
total amount of money in interbank lending transactions exhibits a
growing tendency. Banks are willing to invest their money into
interbank market because it is considered to bear the lowest risk,
so that the lowest obligatory reserves have to be created for the
given credits. Examining the Polish data \cite{a16} we see that
the average equity capital of a standard Polish bank amounts to
about ten percent of all liabilities. The same statistics report
that the average value of assets located in interbank market is
close to fifteen percent. Regulations concerning bankruptcy, which
are common for almost all financial systems, define the bankruptcy
condition as a loss of fifty percent of equity capital.

The numbers quoted show that interbank transactions constitute an
important component of banking assets as well as liabilities
\cite{a14,a15,a16}. Single bankruptcies, through the network of
credit and debt relationships, may seriously threaten the entire
banking system.

In our model vertices on a lattice of the linear size $L$ which
for the simplicity has a regular symmetry (square, cubic or
4d-hypercubic), represent banks. Directed junctions may be
dynamically formed between neighboring nodes. These directed
connections simulate flows of money (interbank transactions, loans
and credits). The paper follows our model of mass bankruptcies
based on the {\it random directed percolation} \cite{a12}. In
comparison to previous work, the major development of the present
model is the implementation of the concept of banking balance,
which when positive, can be invested to make profits, but when
negative, must be refilled to prevent the loss of liquidity. The
next feature distinguishing the proposed model from the previous
one, where a constant concentration of interbank connections was
assumed, is assigning time-dependent weights to directed
connections (i.e. credits) and vertices (i.e. banking capitals).

\begin{figure*} \includegraphics{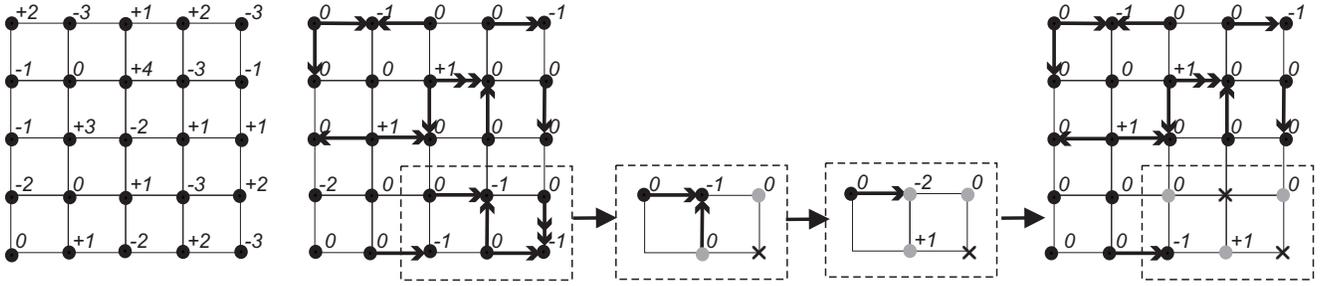}
\caption{\label{fig:1}Dynamical rules in the model. Numbers
describe the unbalanced money inside banks ($d_{ii}$). Black and
grey dots denote uninfected and infected banks respectively, while
crosses correspond to failed banks.}\end{figure*}

The dynamical rules governing the temporal evolution of the model
are as follows: \begin{itemize} \item Initially all banks are
'balanced'. During the following time steps each bank $i$, with
equal probability, experiences shortfalls or surpluses of money
denoted by $\delta_{i}(t)$ (for simplicity $\delta_{i}(t)=\pm 1$).
After time $t$ the banking capital amounts to
$\Theta_{i}(t)=\sum_{\tau=1}^{t}\delta_{i}(\tau)$.
\item At each time step banks with surpluses ($\Theta_{i}>0$) tend to
invest their money, whereas banks with shortfalls ($\Theta_{i}<0$)
want to replenish their resources in order to maintain liquidity.
The best place to realize both aims is the interbank market.
During trading, banking capital is redistributed within the
nearest neighborhood of a bank i.e. interbank financial
transactions (loans and credits) arise. If a $j$-bank owes money
to an $i$-bank we denote the amount of the debt by $d_{ij}>0$. It
follows that $\Theta_{i}=\sum_{<i,j>}d_{ij}$, where
$d_{ij}=-d_{ji}$ denotes the weight of connection between the pair
of neighbors $<i,j>$ and $d_{ii}$ expresses unbalanced money
inside the bank $i$.
\item If a bank has borrowed money in the past period ($\Theta_{i}<0$),
its priority is to repay its creditors. Such a bank when having a
surplus ($\delta_{i}(t)>0$) will intend it for the payment of
debts. On the other hand, a lender-bank ($\Theta_{i}>0$) when
having a short-lived liquidity troubles ($d_{ii}<0$ and $\exists
j$ $d_{ij}>0$) will collect money from its debtors to eliminate a
temporary financial problems.
\item Bank goes bankrupt for either of the two reasons: \begin{itemize}
\item because of the loss of solvency, when its capital falls below the
'solvency threshold' $\Theta_{i}\leq \Theta_{s}$, or
\item because of the loss of liquidity, when unbalanced money
inside a bank exceeds the 'liquidity threshold' $d_{ii}\leq
\Theta_{l}$ where $\Theta_{s}< \Theta_{l}< 0$.\end{itemize}
\item Lender-banks that have given credits to a bankrupt lose their
money. We assume that these banks become 'infected' and they
attempt to secure their positions by collecting their money from
the debtors. The borrower-banks that have to repay incurred debts
are also assumed as 'infected'. These banks propagate infection to
their further uninfected lenders. The effect of contagion
transmission by infected borrower-bank is related to the situation
when a lender having a contact with an infected debtor wants to
protect itself against losses and tries to collect given credits.
In this way, a bankruptcy of a single bank may initiate an
avalanche of bank failures similarly to the {\it domino effect}
well known in social and economic modeling.
\item When the avalanche stops, new balanced banks ($\Theta_{i}=0$) are
created to replace the failed ones.
\end{itemize}
The above dynamical rules of the model become comprehensible after
examining Fig. \ref{fig:1} which presents an example of a banking
network on the square lattice of size $5\times5$.
 For simplicity
we denote a bank located in the row $k$ and in the column $l$ by
the subscript $i=(k-1)*5+l$. The figure was generated under the
following assumptions: $\delta_{i}=\pm 1$, $\Theta_{s}=-4$,
$\Theta_{l}=-2$. A random set of variables $\Theta_{i}$ as an
initial condition was assumed. There is no interbank market at the
first stage of the figure. Effects of interbank trading appear at
the second stage. Arrows with single or double arrowheads
correspond to $d_{ij}=\pm 1$ or $\pm 2$ respectively. After
trading, the interbank market is almost balanced. Only one failure
resulting from the loss of liquidity happened (the bank $i=16$).
However because there were no commitments of this bank to its
neighbours thus this bank failure did  not lead to further network
contagion. Let us now assume that the last bank ($i=25$) faces a
shortfall ($\delta_{25}=-1$) in the next time step (our model is
governed by asynchronous dynamics). This forces its immediate
bankruptcy due to the loss of solvency ($\Theta_{25}=-4$). The
failure infects two other banks with numbers $20$ and $24$
respectively. Since the $24^{th}$ bank tries to collect the credit
given to $19^{th}$ bank it forces its failure according to the
liquidity condition. The second bankruptcy infects the $18^{th}$
bank. In this way the contagion process originating from the
$25^{th}$ node triggers off a small avalanche of two failures
within the cluster of five 'infected' / failed banks.

\section{Computer simulations and results}
\begin{figure}
\includegraphics[angle=-90,scale=0.5]{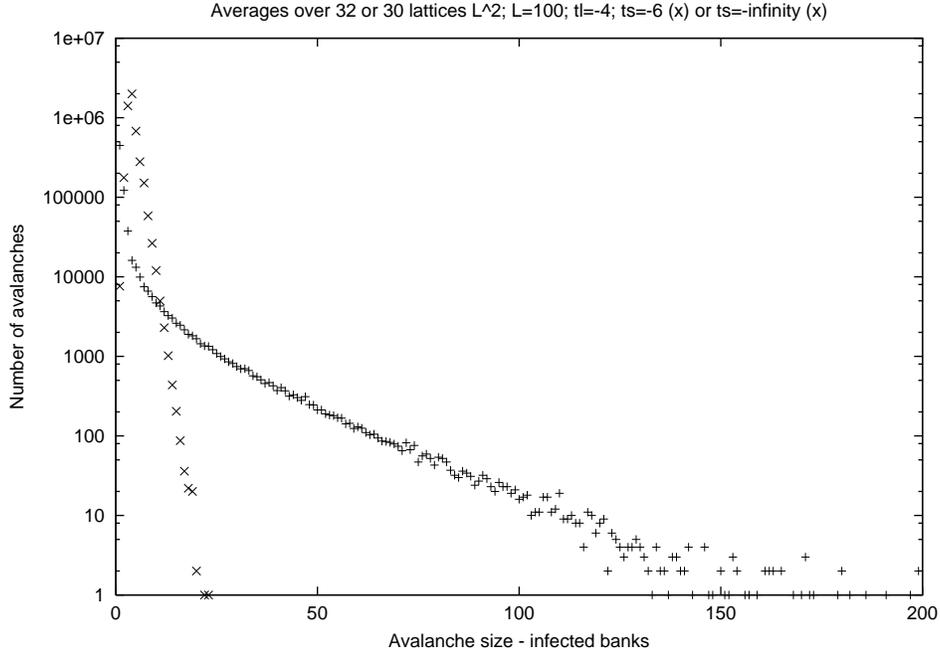}
\caption{Distributions of
avalanches in square lattices} 
\label{fig:2}
\end{figure}

\begin{figure}
\includegraphics[angle=-90,scale=0.5]{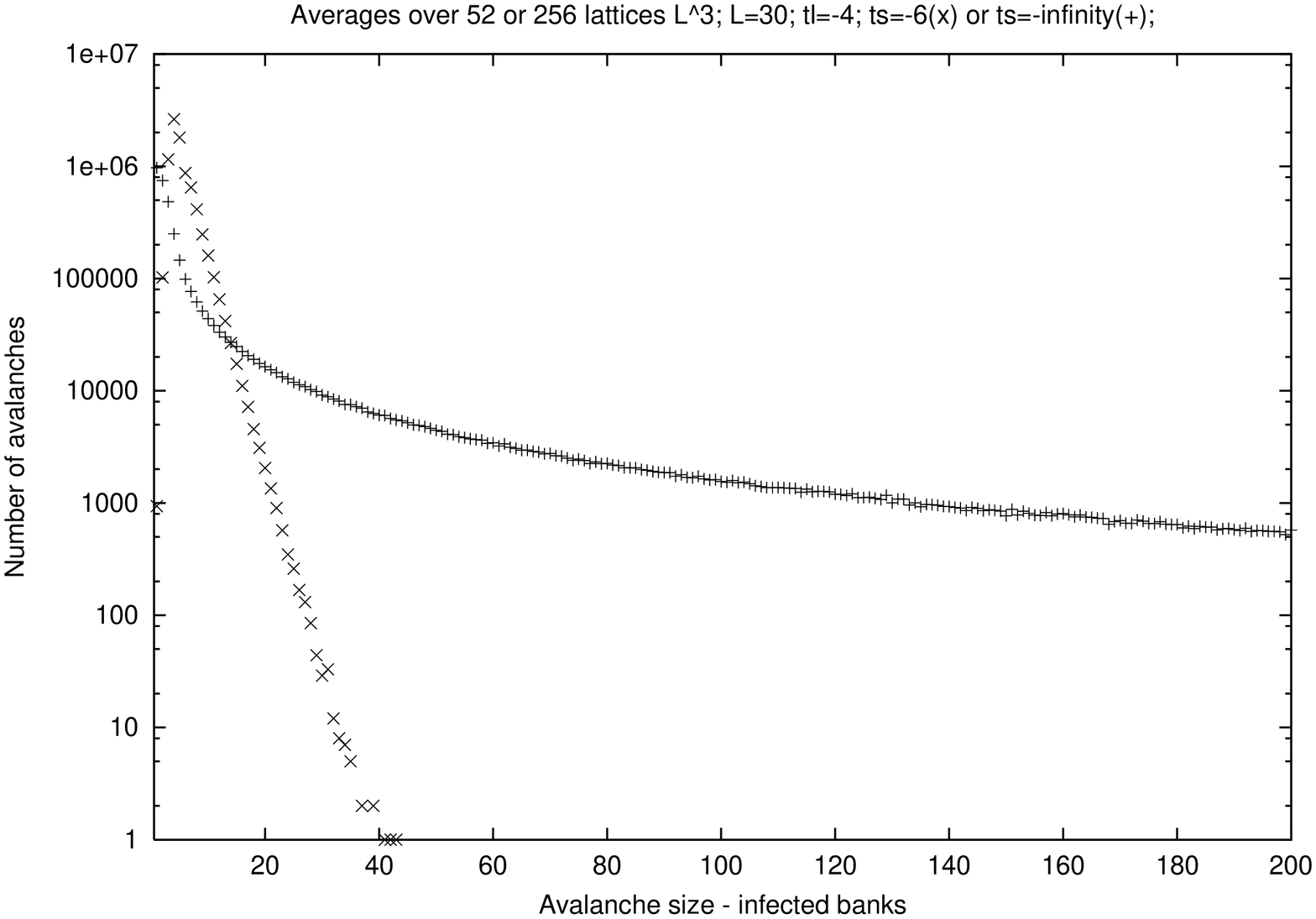}
\includegraphics[angle=-90,scale=0.5]{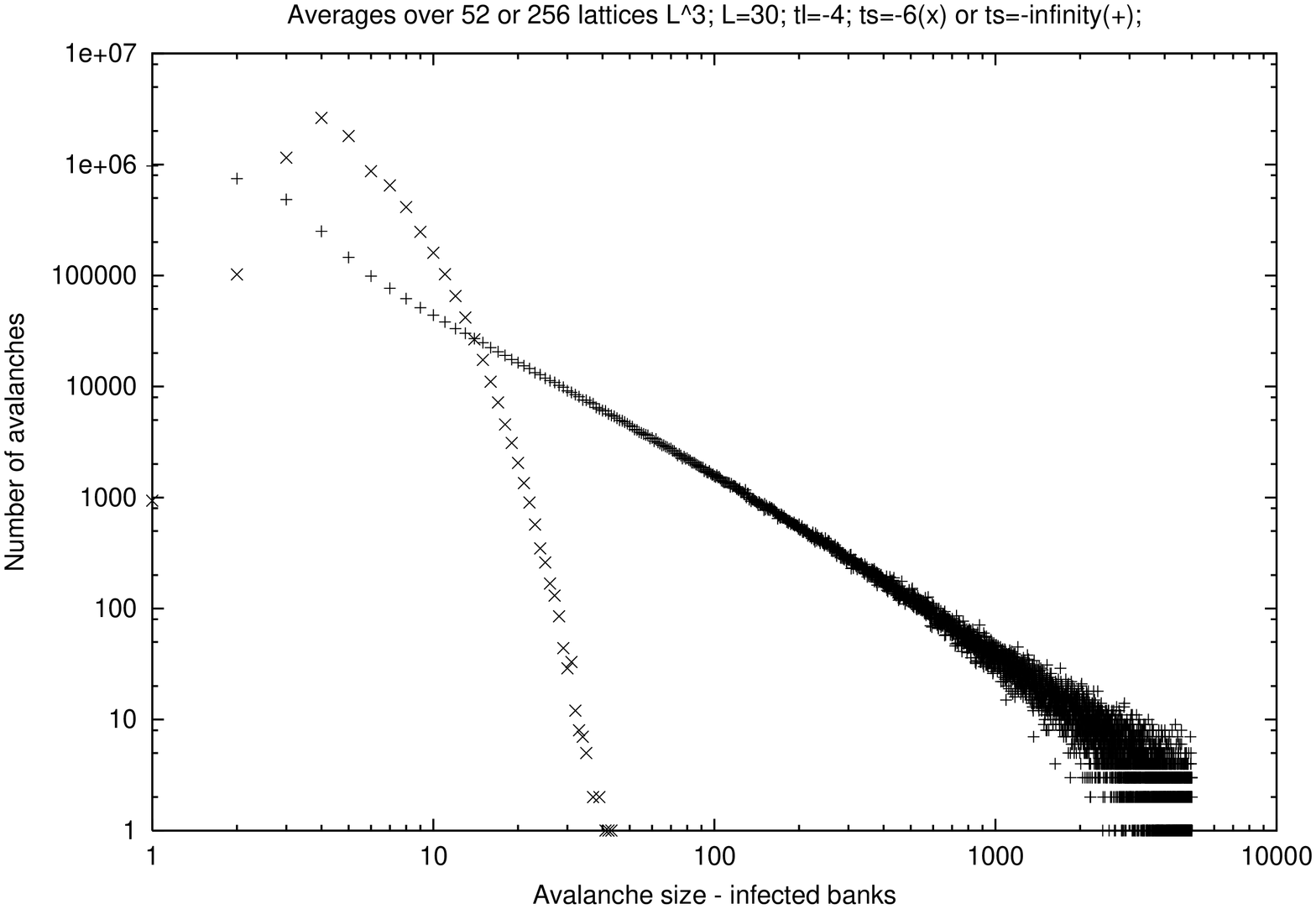}
\caption{Distributions of
contagion area in 3d-cubic lattices. Top: semi-logarithmic plot, bottom: 
logarithmic plot}
\label{fig:3}
\end{figure}

\begin{figure}
\includegraphics[angle=-90,scale=0.5]{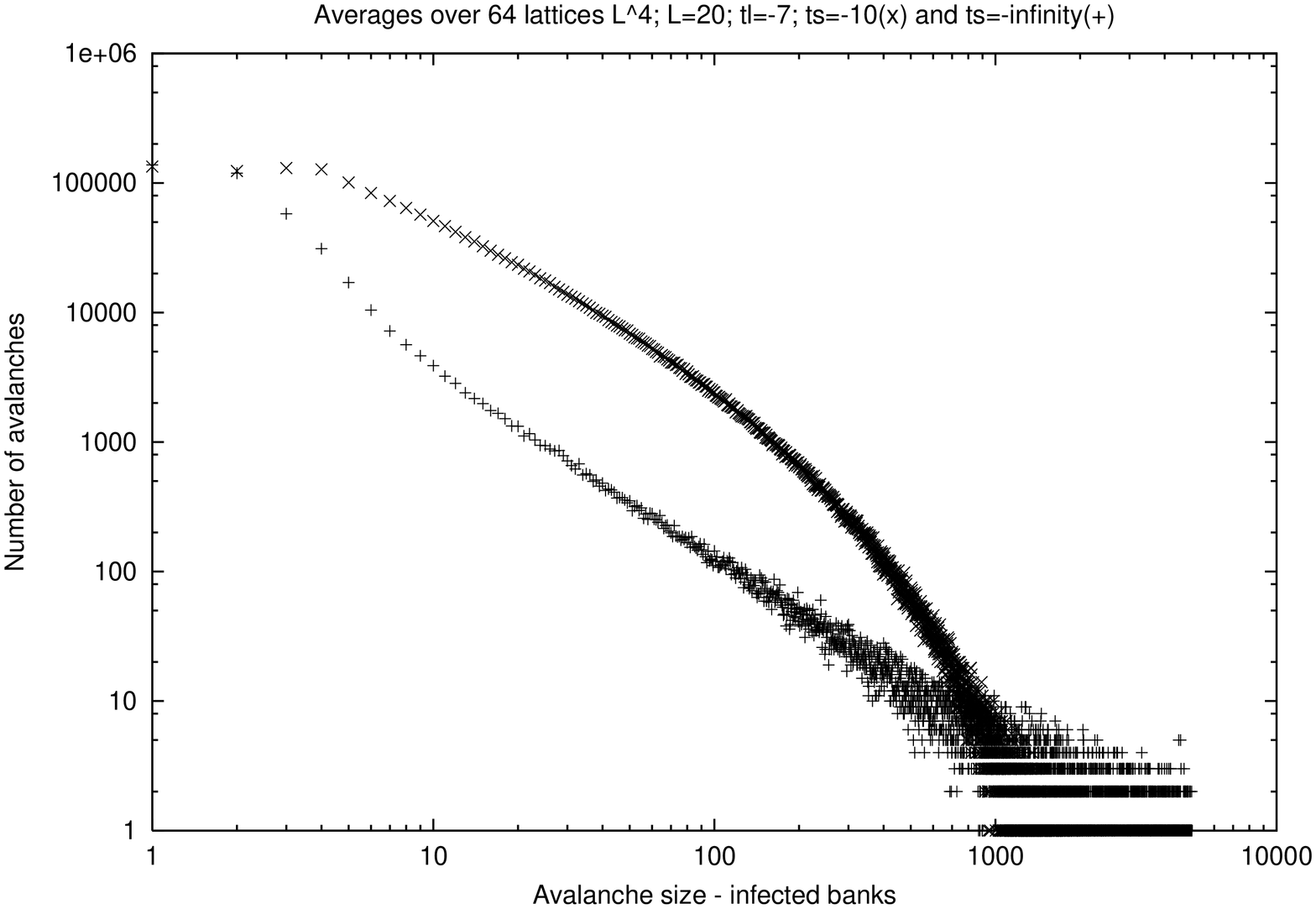}
\caption{Distributions of
avalanches in 4d-hypercubic lattices}
\label{fig:4}
\end{figure}

We investigated effects of a single bank failure on the nucleation
of contagion phase in the banking network. The simulations were
done for square, cubic and 4d-hypercubic lattices. Although one
could expect similar results for all dimensions, we observe a
noticeable difference in behavior. The distribution of infection
area in the networks based on square lattice is exponential
(Fig.\ref{fig:2}), whereas the same dynamical rules in
four-dimensional hypercubic lattices generate power law
distributions (Fig.\ref{fig:4}). The fact that the contagion in
two-dimensional systems is characterized by exponentially decaying
distribution of stopping times  indicates {\it subcritical}
behavior. For 4d-hypercubic lattices, the contagion process may
continue infinitely as the result of long-range spatial and
temporal correlations between the participants of the interbank
market. Three-dimensional cubic lattices described by different
combinations of the system parameters ($\Theta_{s}$ and
$\Theta_{l}$) exhibit a continuous transition from sub-critical to
critical behavior of the model (Fig.\ref{fig:3}).

These differences can be explained as follows. The expansion of
the infection area corresponds to avalanches of random directed
percolation. On square lattices, the neighborhood of each vertex
consists of only four vertices. In terms of our model, it means
that the average number of potential financial partners of a bank
(neighbors whose banking capital $\Theta_{i}$ has an opposite
sign) equals to two. Given the above considerations, we conclude
that the average concentration of interbank connections $q$ should
not exceed fifty percent of all possible links in the lattice
($q\leq 0.5$). The random directed percolation threshold satisfies
$p_{c}^{2d}=1.0$ on square lattices \cite{a12}. Comparing two
parameters $q<p_{c}^{2d}$ makes it clear that two-dimensional
banking networks always remain in the sub-critical region of the
percolation phase transition. On 4d-hypercubic lattices, the
threshold amounts to $p_{c}^{4d}\approx 0.3202$ while $q\leq 0.5$.
For this reason the four-dimensional interbank market driven by
threshold dynamics of the participants can balance at the edge of
percolation phase transition in a self-organized critical state almost
regardless of the system parameters values.  On the other hand the
three-dimensional percolation threshold $p_{c}^{3d}\approx 0.4976$
is only slightly smaller than the maximum concentration of
interbank connections $q_{max}=0.5$. Tuning the solvency threshold
with the fixed liquidity parameter shows that although the
3d-model tends to the self-organized critical dynamics,  for
certain combinations of system parameters ($\Theta_{s}$ and
$\Theta_{l}$) the double security system consisting of liquidity
and solvency monitoring keeps it in the sub-critical region
(Fig.\ref{fig:3}).

We interpret the power law (free-scale) distributions of failure
sizes in 3d-cubic and 4d-hypercubic lattices as a symptom of self-organized
critical behavior of the model. The slopes of the histograms
approach the same value $\tau^{3d}_{SOC}=\tau^{4d}_{SOC}\approx1.50\pm0.05$
in both cases. Despite these reasonable power-laws indicating SOC dynamics, 
some non-equilibrium effects in the model are not yet understood.   
Large systems in higher dimensions are particularly sensitive to these effects.

To understand the SOC phenomenon on the microscopic level one
needs to consider the dynamical character of the system parameter
$q$. In fact there are two competing processes that increase or
decrease the mean number of interbank connections $q$.

\begin{description} \item[A.] Due to the random character of the stochastic
variable $\delta_{i}(t)$ and the random selection of a financial
partners from the nearest neighbors, the number of interbank
connections $q$ is growing up. \item[B.] Due to collective
bankruptcies the number of interbank connections $q$ is falling
down because new banks that replace bankrupts are not involved in
the interbank market at the beginning of their activity.
\end{description}
If the value of $q$ is below the percolation threshold then the
mean size of contagion cluster is small and the process $A$
overcomes the process $B$. It follows that in the course of  time
the number of interbank connections $q$ is growing up driving the
system to the critical state. If the value of $q$ exceeds a
critical value (the percolation threshold) then a large contagion
cluster spreading all over the system can appear. As a result a
lot of banks that were involved in the network of interbank
transactions face bankruptcy. The process $B$ overcomes the
process $A$ and the parameter $q$ is diminishing.

\section{Discussion}
The model presented here relates the phenomena of collective
bankruptcies in banking networks to the self-organized dynamics
driving the system towards the critical point. Although a similar
effect of power law scaling of collective bank failures was
partially discussed in \cite{a12,a13}, the previous approaches
were related to externally controlled critical parameters and did
not translate into terms of self-organized criticality. In the
present model, due to the inclusion of more realistic assumptions
concerning the interbank market the connectivity parameter $q$ can
oscillate at the edge of the phase transition. At this point we
would like to stress, that however we abstract from details of
banking mechanisms and encompass all of them in only one parameter
(the unit balance $\delta_{i}(t)$) thanks to this simplification,
the model becomes more comprehensible and gains universality.

One could imagine a number of further extensions of the model.
Since the real banking networks do not possess any regular
symmetry it would be instructive to analyze the model from the
point of view of the random graph architecture. The model
introduced here bases on the assumption of a fixed number of
nearest neighbors. The above situation is not observed in reality
but as our results emphases, the artificial lowering of the
dimensionality of the network keeps the model in the subcritical
state. Since in the four-dimensional network, the double security
system consisting of liquidity and solvency monitoring does not
bring any effective control of the whole network security, the
control should be established over the number of debtors or
amounts of the credits.

\begin{acknowledgments}
We are grateful to Prof. Dietrich Stauffer for the critical reading of
this manuscript and for making the Crey-T3E from J\"{u}lich Supercomputer Center
available to numerical tests. Two of us (AA and JH) are thankful to Prof. Dirk
Helbing for his hospitality during the stay in Dresden. The work
has been in part supported by the ALTANA AG due to the Herbert
Quandt-Programm and by the DAAD.
\end{acknowledgments}

\end{document}